# Micro-electroforming Metallic Bipolar Electrodes for Mini-DMFC Stacks

R. F. Shyu[1], H. Yang[2], J.-H. Lee[2]

[1]Department of Mechanical Manufacturing Engineering, National Formosa University, Yunlin, Taiwan 632

[2]Institute of Precision Engineering, National Chung Hsing University, Taiwan 402

*Abstract-* This paper describes the development of metallic bipolar plate fabrication using micro-electroforming process for mini-DMFC (direct methanol fuel cell) stacks. Ultraviolet (UV) lithography was used to define micro-fluidic channels using a photomask and exposure process. Micro-fluidic channels mold with 300 μm thick and 500 μm wide were firstly fabricated in a negative photoresist onto a stainless steel plate. Copper micro-electroforming was used to replicate the micro-fluidic channels mold. Following by sputtering silver (Ag) with 1.2μm thick, the metallic bipolar plates were completed. The silver layer is used for corrosive resistance. The completed mini-DMFC stack is a 2x2 $cm^2$ fuel cell stack including a 1.5x1.5 $cm^2$ MEA (membrane electrode assembly). Several MEAs were assembly into mini-DMFC stacks using the completed metallic bipolar plates. All test results showed the metallic bipolar plates suitable for mini-DMFC stacks. The maximum output power density is 9.3mW/$cm^2$ and current density is 100 mA/$cm^2$ when using 8 vol. % methanol as fuel and operated at temperature 30°C. The output power result is similar to other reports by using conventional graphite bipolar plates. However, conventional graphite bipolar plates have certain difficulty to be machined to such micro-fluidic channels. The proposed micro-electroforming metallic bipolar plates are feasible to miniaturize DMFC stacks for further portable 3C applications.

## I. INTRODUCTION

Global warming effect results in serious environmental problems. Fossil energy consumption generated lots of carbon dioxide and related compound is the major reason. Clean energy with high conversion efficiency and low pollution effect is human's common objective. Fuel cell is an electrical power generator, it converts chemical energy to electrical power. There are two major fuel cells including PEMFC (proton exchange membrane fuel cell) and DMFC. Both systems need a FC stack as the reaction site. The key components include proton exchange membrane, gas diffusion layer (electrode), catalyst, and bipolar plate. The combination of proton exchange membrane, catalyst and gas diffusion layers is called MEA (membrane electrode assembly). Hydrogen fuel with catalyst help reacting with oxygen generates electricity and water. Bipolar plates are responsible for hydrogen and water flow in the FC stack.

Many conductive materials are suitable for mini-DMFC bipolar plates, silicon and stainless steel ate two major items [1]. Jiang et al used (100) silicon as the substrate and etched microchannels as bipolar plates [2]. A reactive area of 1.3×1.1$cm^2$ MEA was applied and generated the maximum output power density 2.31 mW/$cm^2$. Methanol solution (1 M volume concentration) was operated at room temperature. Though silicon bipolar plates can be fabricated for micro-DMFC stacks, the brittle problem is easy to make cracks in assembly and results in failure [1]. Stainless steel provides low cost and high strength advantages compare to silicon. The skillful metal work can realize stainless steel for commercial applications [3]. Yang et al used stainless steel (SS316) as bipolar plate material for micro-DMFC stacks. A MEA with 5 $cm^2$ was assembly to a single DMFC. The maximum output power was 13mW/$cm^2$ when using 6 vol. % methanol solution and operated at temperature 31°C [4].

Different microchannel flow patterns have direct influence on fuel cell stack performance. Serpentine flow pattern has the best result based on practical test compared to other patterns. Bipolar plates with serpentine flow pattern contributed a better $CO_2$ exhaust on anode side. The further study also indicated the generated output efficiency using serpentine flow pattern is suitable for micro-DMFC stacks [5]. Microchannel profile related to liquid flow pattern is the other concerned issue. Microchannel with 580μm wide for mini-DMFC has a better performance than other sizes [6]. Different depths were studied and concluded mcirochannel with 300 μm deep suitable for the optimum efficiency [7]. Based on previous studies, many researchers suggested that fluidic channel with 300 μm in depth and 500 μm in width provides an efficient flow.

The experiment was to design such pattern in the mask. UV-LIGA process was applied to fabricate such pattern for bipolar plates with low electrical resistance. Stainless steel (SS316) with 1.5 mm thick was used as the substrate. The negative photoresist (JSR THB-120N) with 300 μm thick was triple-spun coating on the substrate. UV exposure and development then generated the fluidic channel mold in photoresist. Copper electroforming then filled the mold onto the substrate. The dimensional measurement is required to confirm the geometry accuracy by using an optical 3D profile. Electrical performance measurement of mini-DMFC to verify the experimental bipolar plates is measured by a commercial MEA and test system.

## II. EXPERIMENTAL PROCEDURE

Metallic bipolar plates with different patterns for microchannel flow were studied. Previous studies showed that the optimum microchannel with width 500μm and depth 300μm provides the best efficiency. Pattern design on the mask is shown in Fig. 1. Serpentine (left diagram) and interdigital (right diagram) flow patterns were designed as shown. Desired patterns are transferred from the designed mask in the lithographic process. In this experiment, a plastic mask was fabricated using laser writing onto a PET (Polyethylene terephthalate) used for PCBs (print circuit boards). The stainless steel (SS316) substrate (Fig. 2) was then spun with a layer of negative photoresist (JSR THB-120N) 100μm thick. The spin condition was 150rpm for 15 seconds. Prebaking in a convection oven at 100°C for 3 minutes is a required procedure. This removes the excess solvent from the photoresist and produces a slightly hardened photoresist surface. Repeat the coating and prebaking process in three times resulted in multi-layer thickness 300 μm. The mask was not stuck onto the substrate.

The sample was exposed through the plastic mask using a UV mask aligner (EVG620). This aligner had soft, hard contact or proximity exposure modes with NUV (near ultra-violet) wavelength 350-450nm and lamp power range from 200-500 W. The exposure time was 180 seconds and developing time for 40 minutes. The hardbake temperature was 120°C for 3 minutes by the hot plate. Commercial cooper sulfite electrolyte was used to electroform metallic plates. The specimen with microchannel flow pattern on the stainless steel substrate was immersed into the electrolyte. Cooper atoms were deposited into resist mold from the substrate. The input current density was 5ASD and the growth rate was 1 μm/min in theory. The plating elapsed for 5 hours to obtain microchannel thickness 300μm in cooper. Sodium hydroxide solution 150 ml containing 3.8 g NaOH was heated to 60°C. Photoresist mold was detached from the substrate after 15 minutes immersion. After cleaning with DI water and spray nitrogen air, metallic cooper microchannel bipolar plates were completed. After stripping the photoresist and sputtering silver with 1.2 μm, the metallic bipolar plate with surface resistance 284 mΩ was completed. The completed metallic bipolar plates for mini-DMFC stacks are shown in Fig. 3.

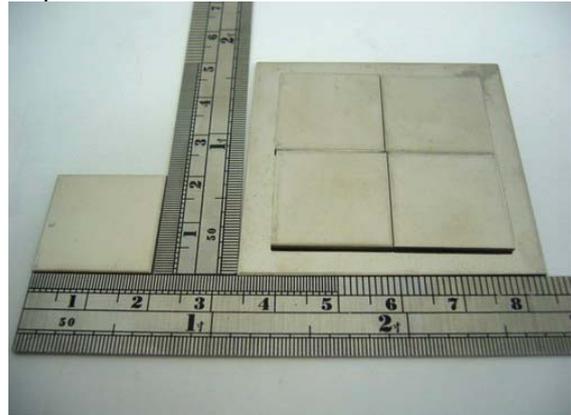

Fig. 2. Stainless steel (SS316) substrates as the plating base material, each piece is 2x2 cm$^2$.

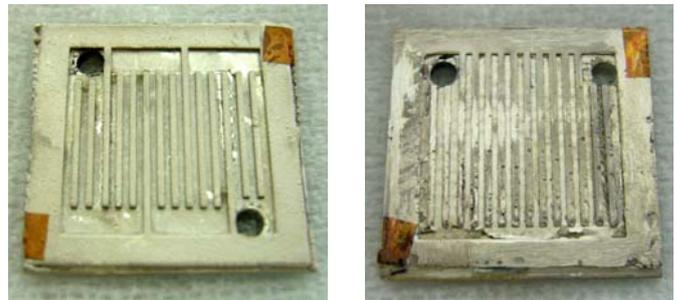

Fig. 3. Completed metallic bipolar plates for mini-DMFC stack.

## III. EXPERIMENTAL RESULTS

### A. Dimensional measurement

Serpentine microchannel profiles of bipolar plates were measured using a 3D surface profiler to measure their widths and depths. As shown in Fig. 4(a), the measured width is 498μm and the depth is 350μm. The three dimensional profile is also shown in Fig. 4(b). Rough surface came from large grains of copper deposition. That can be further modified in electroforming conditions such decreasing growth rate and adding some additives for grain refining. Interdigital microchannel pattern of bipolar plates was measured as shown in Fig. 5. The dimensional measurement results are close to the design size. There is about 10% dimensional error existed.

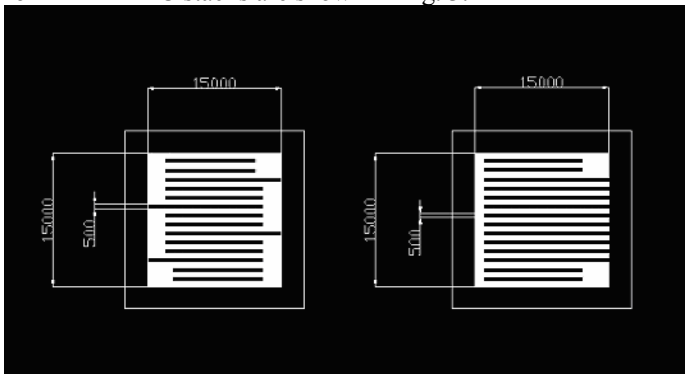

Fig. 1. Schematic diagrams of microchannel flow pattern on the mask.

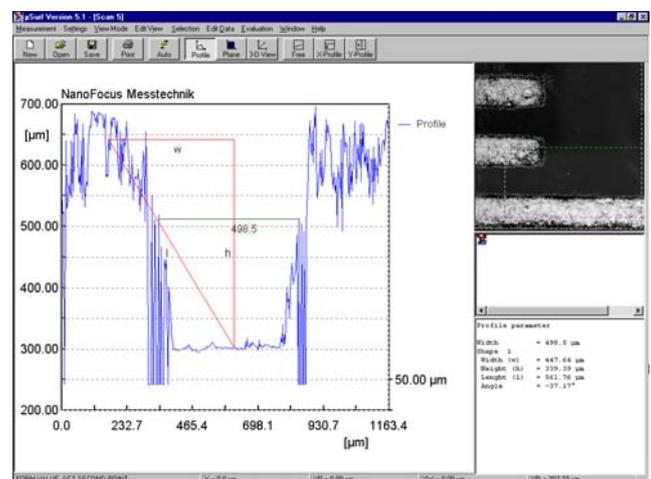

(a)

platform. The maximum output power density 12 mW/cm$^2$ with the current density 50 mA/cm$^2$ was measured. That's a reference to verify the experimental metallic bipolar plates since the commercial bipolar plate is made of graphite. Serpentine and interdigital microchannel patterns of metallic bipolar plates in mini-DMFC were assembly and measured their performance characteristics. Fig. 9 shows the performance characteristics of serpentine microchannel bipolar plates, its output performance is similar to the commercial product. Interdigital microchannel bipolar plates have the poor performance as shown in Fig. 10. Once again showing serpentine flow pattern is more suitable for DMFC stacks. Fig. 11 shows the functional test of mini-DMFC by using a mini-fan as the load. The output voltage 0.37 V is shown to drive the mini-fan running.

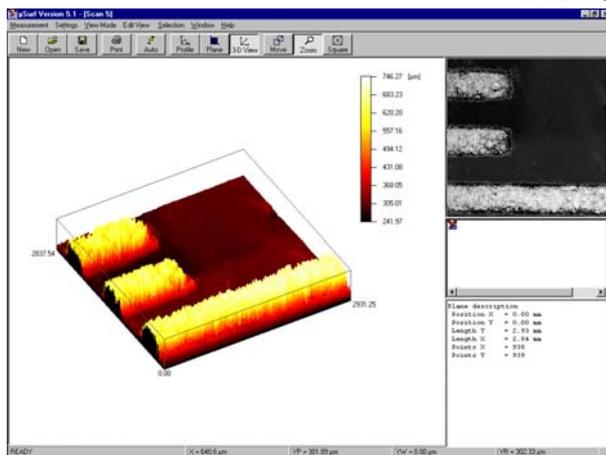
(b)
Fig. 4. Measurement results of serpentine microchannel pattern bipolar plate; (a) cross-sectional view and (b) 3D morphology.

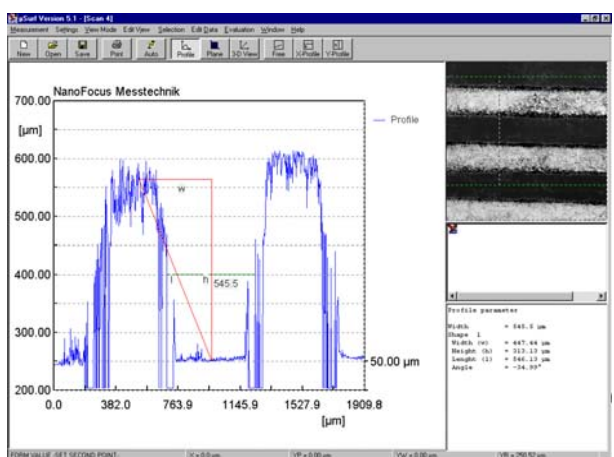
(a)

(b)
Fig. 5. Measurement results of interdigital microchannel pattern bipolar plate; (a) cross-sectional view and (b) 3D morphology.

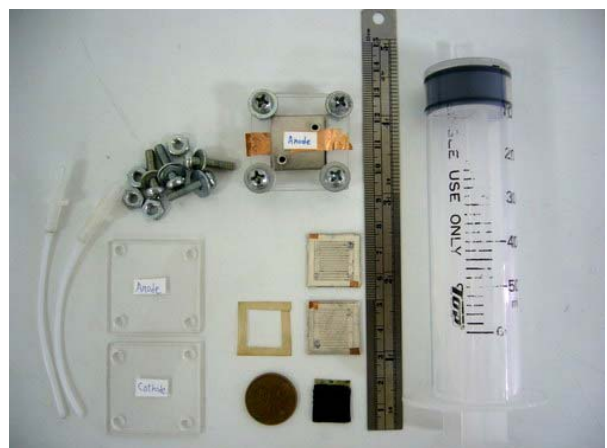
Fig. 6. Illustration of mini-DMFC stack parts.

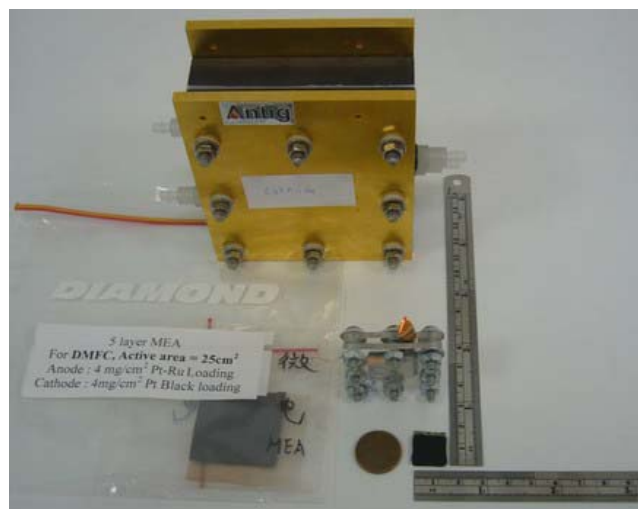
Fig. 7. Comparison photos of conventional DMFC stack and mini-DMFC stack.

*B. Mini-DMFC stack assembly*

The mini-DMFC stack includes a 1.5x1.5cm$^2$ MEA from Micropower Tech, metallic bipolar plates from this experiment, and other fixtures as shown in Fig. 6. To understand the commercial DMFC performance using Micropower Tech's MEA, Fig. 8 shows its performance characteristics. Test equipment was the Antig's DMFC tes

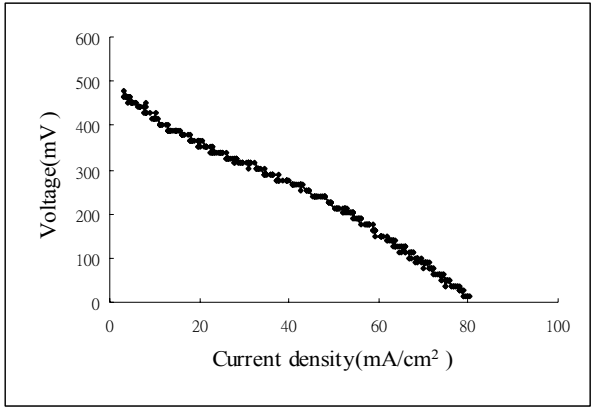

(a)

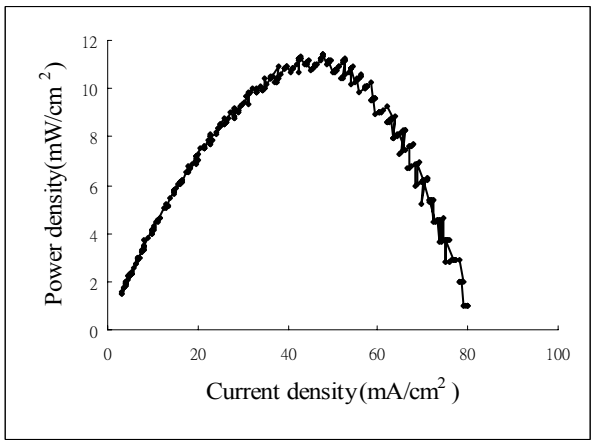

(b)

Fig. 8. Conventional DMFC performance characteristics by using MEA (3.5×3.5cm$^2$) from Micropower Tech.; (a) I-V curve and (b) output power.

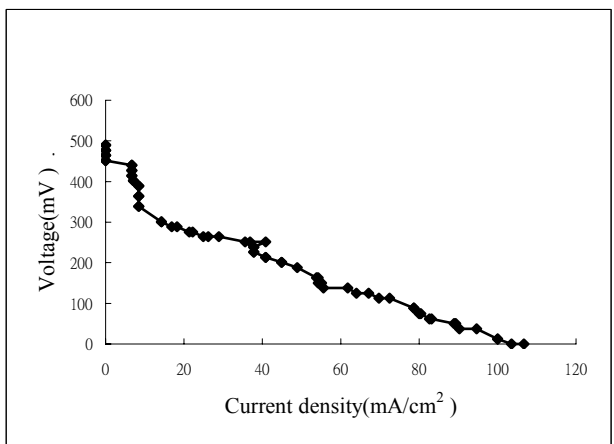

(a)

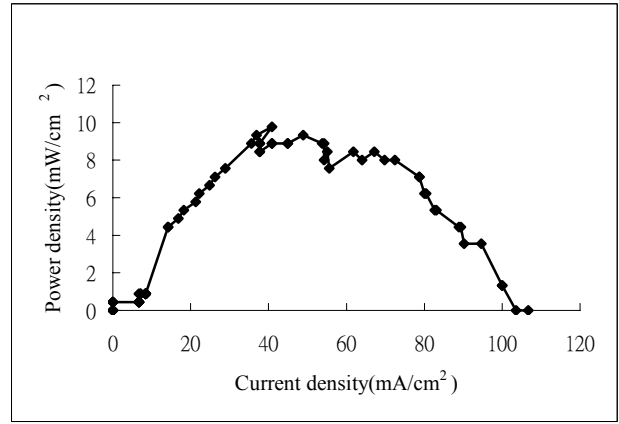

Fig. 9. Mini DMFC performance characteristics using serpentine microchannel pattern bipolar plates; (a) I-V curve and (b) output power.

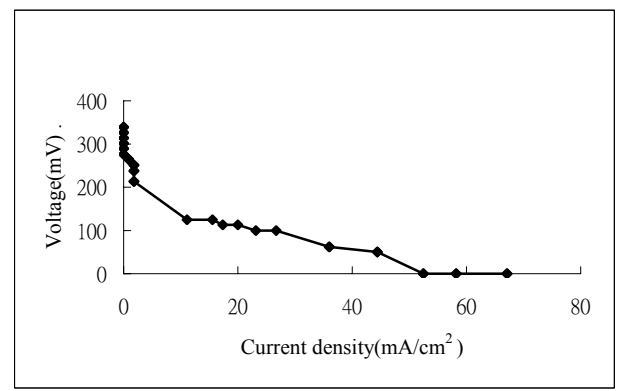

(a)

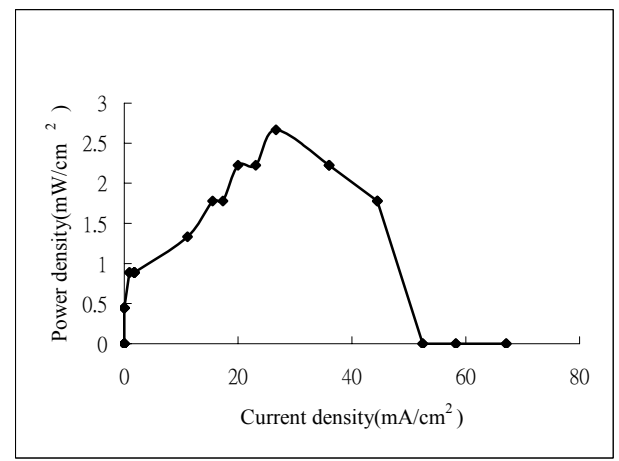

(b)

Fig. 10. Mini DMFC performance characteristics using interdigital microchannel pattern bipolar plates; (a) I-V curve and (b) output power.

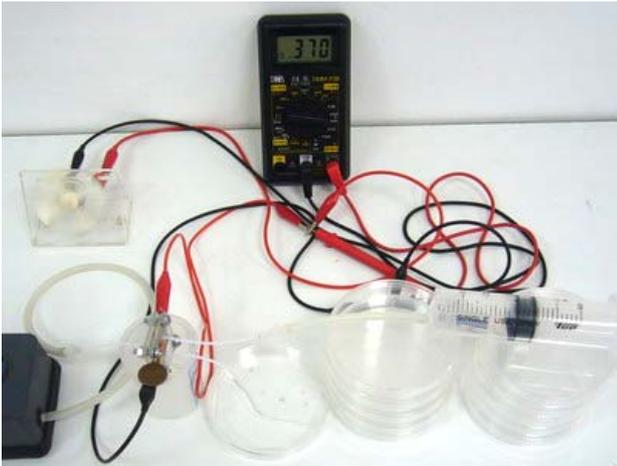

Fig. 11. Functional test of mini DMFC stack with output voltage 0.37 V.

## IV. CONCLUSION

Commercial MEAs with the fabricated bipolar plates were assembly into a 2x2 $cm^2$ stack. As found, the output power mainly dominated by MEA. MEA is the most important factor effect on the output power, though fluidic channel types on bipolar plates have a little effect on the stack performance. The completed mini-DMFC stack has the maximum power 9.33mW/$cm^2$ was measured. The power density is similar to macro-FC stacks. It proves the proposed micro-electroforming bipolar plates feasible to mini-DMFC stacks. The further applications trend to apply the mini-DMFC stacks for portable electronic products.

ACKNOWLEDGMENT

This research is supported by the National Science Council of Taiwan under the grand number NSC96-2221-E-005-069-MY3.